\documentclass[aps,pre,longbibliography,onecolumn,superscriptaddress,showpacs]{revtex4-2}

\usepackage{CJK}

\usepackage{graphicx}
\usepackage{hyperref}
\usepackage{amsmath}
\usepackage{dcolumn}
\usepackage{bm}
\usepackage[normalem]{ulem}
\usepackage{xcolor}

\usepackage{epstopdf}

\newcommand{\bew}{\begin{widetext}}
\newcommand{\ew}{\end{widetext}}

\newcommand{\ii}{{\rm i}}

\newcommand{\bq}{\mathbf{q}}
\newcommand{\bv}{\mathbf{v}}

\newcommand{\br}{\mathbf{r}}

\newcommand{\bff}{\mathbf{f}}
\newcommand{\bu}{\mathbf{u}}

\newcommand{\bw}{\mathbf{w}}
\newcommand{\bbr}{\mathbf{r}}

\newcommand{\bk}{\mathbf{k}}

\newcommand{\beq}{\begin{equation}}
\newcommand{\eeq}{\end{equation}}
\newcommand{\beqn}{\begin{eqnarray}}
\newcommand{\eeqn}{\end{eqnarray}}
\newcommand{\pp}{\partial}
\newcommand{\dd}{{\rm d}}
\newcommand{\ee}{{\rm e}}

\newcommand{\la}{\langle}

\newcommand{\ra}{\rangle}

\newcommand{\vnab}{{\bf \nabla}}

\newcommand{\ctea}{\color{teal}}

\begin{document}
\title{Incompressible polar active fluids  with quenched disorder in dimensions $d> 2$}
\author{Leiming Chen}
\email{leiming@cumt.edu.cn}
\address{School of Material Science and Physics, China University of Mining and Technology, Xuzhou Jiangsu, 221116, P. R. China}
\author{Chiu Fan Lee}
\email{c.lee@imperial.ac.uk}
\address{Department of Bioengineering, Imperial College London, South Kensington Campus, London SW7 2AZ, U.K.}
\author{Ananyo Maitra}
\email{nyomaitra07@gmail.com}
\address{Laboratoire de Physique Th\'eorique et Mod\'elisation, CNRS UMR 8089,
	CY Cergy Paris Universit\'e, F-95302 Cergy-Pontoise Cedex, France}
\author{John Toner}
\email{jjt@uoregon.edu}
\affiliation{Department of Physics and Institute of Theoretical
Science, University of Oregon, Eugene, OR $97403^1$}
\affiliation{Max Planck Institute for the Physics of Complex Systems, N\"othnitzer Str. 38, 01187 Dresden, Germany}
\date{\today}

	\begin{abstract}
	We  present a   hydrodynamic  theory  of incompressible polar active fluids with quenched disorder. This theory shows that such fluids can overcome the disruption caused by the quenched disorder and move coherently, in the sense of having a non-zero mean velocity in the hydrodynamic limit.
 However, the scaling behavior of this class of active systems can{\it not} be described by linearized hydrodynamics in spatial dimensions between 2 and 5. Nonetheless, we obtain the   exact dimension-dependent scaling exponents in these dimensions.
	\end{abstract}
\maketitle


One of the most  important themes of condensed matter physics is the competition between order and disorder. One of the most powerful results on this topic is the Mermin-Wagner-Hohenberg theorem \cite{MW1, MW2}, which states that {\it equilibrium} systems cannot spontaneously break a continuous symmetry in spatial dimensions $d\le2$ at non-zero temperature.
Much of the current interest in ``active matter" is stimulated by the discovery \cite{vicsek_prl95, toner_prl95, toner_pre98, Malthus}
that non-equilibrium ``movers" {\it can}
spontaneously break a continuous symmetry (rotation invariance) in the presence of noise even in $d=2$, by ``flocking"; that is, moving coherently with
a  non-zero spatially averaged velocity $\left<{\bf v} ({\bf r}, t) \right>\ne \bf 0$.

In equilibrium systems,
even {\it arbitrarily weak} quenched (i.e., static) random fields  destroy long-ranged ferromagnetic order in all spatial dimensions $d\le4$ \cite{Harris, Geoff, Aharonyrandom,Dfisher}. This
raises the question:
can an ordered polar active fluid form  when quenched disorder is present?

This question is relevant  to biology. For example, it arises for cells moving through an extracellular polymerised matrix. Such matrices inevitably contain local random spatial heterogeneities which are fixed on the experimentally relevant  timescale,  and therefore, quenched \cite{grinnell_annrev10}.
Given its experimental relevance, the effects of quenched disorder on active matter has therefore received much attention recently \cite{toner_prl18, toner_pre18,Duan21, Dor21, Ro21, Peruani1, Peruani2, Peruani3, Chardac, Bartolo1}

In this Letter, we investigate a heretofore unconsidered situation: {\it incompressible} polar active fluids with quenched disorder.  This is relevant in the cellular context when motile cell clusters maintain a constant density as  move. Incompressibility could arise either due to cell-cell avoidance by long-distance sensing through fast-diffusing signaling molecules, or  because of steric interactions in the high packing limit \cite{chen_njp15}. Our key result is
that, in all spatial dimensions  $d>2$, an ordered polar active fluid phase
survives in the presence of a finite amount of quenched disorder. Furthermore, we find that for $2<d<5$, there is a {\it breakdown of linearized hydrodynamics}, just as there is in simple fluids \cite{forster_pra77} for $d\le2$, and flocks without quenched disorder for $d\le4$ \cite{toner_prl95, toner_pre98}. That is, the spatio-temporal scaling of fluctuations in these systems is {\it not} correctly {given}
by a linear theory, due to strong non-linear coupling between large fluctuations. Nonetheless, there {\it is} universal scaling of correlations in this range of spatial dimensions, and we have been able to determine its scaling exponents {\it exactly}.

In previous papers \cite{us}, we have shown  that incompressible polar active fluids retain long-range order even in $d=2$ in the presence of quenched disorder. Since the effect of fluctuations is expected to reduce with increasing dimensionality, this would seem to directly imply long-range order for all $d>2$ as well. However, the incompressible flock in $d=2$ is qualitatively distinct from that in higher dimensions \cite{chen_natcomm16, chen_njp18} since it lacks a true ``soft'' or hydrodynamic mode for most directions of wavevector
because incompressibility constrains the dynamics to a \emph{much} greater degree in $d=2$. As a result, the findings in Ref.~\cite{us} do not automatically  imply  long-range order in $d>2$. Our conclusion here that there {\it is} long range order in all $d>2$
is therefore nontrivial and new.

In the following, we will first present a hydrodynamic theory of incompressible polar active fluids with {\it both}  annealed disorder (which represents endogenous fluctuations due to, e.g., errors made by a motile agent while attempting to follow its neighbors \cite{vicsek_prl95}) and quenched disorder.
We then apply a dynamic renormalization group (DRG)  analysis to obtain the exponents that fully characterize the scaling behavior of the system in the moving phase. Specifically, choosing our coordinates so that the   $x$-axis is along the mean velocity $\la \bv \ra $ of the flock  (i.e., $\la \bv \ra = v_0 \hat{\bf x}$),  and
defining the fluctuation   $\bu(\br, t)$ of the velocity at the point $\br$ at time $t$ away from this mean velocty via $\bu(\br, t) = \bv(\br, t)-v_0 \hat{\bf x}$,
we find that the two point correlations
$\la \bu(\br, t) \cdot \bu(0, \mathbf 0)\ra$ of these fluctuations is of the form
\beqn
\la \bu(\br, t) \cdot \bu(0, \mathbf 0)\ra&=&
r_{_\perp}^{2\chi}G_{_{Q}}\left(|x|\over r_{_\perp}^{\zeta}\right)
\nonumber\\
&+&r_{_\perp}^{2\chi'}G_{_{A}}\left({|x-\gamma t|\over r_{_\perp}^{\zeta'}}, {|t|\over r_{_\perp}^{z'}}\right)\, ,
\label{3Dcorrelation1}
\eeqn
where $G_{_Q}$ and $G_{_A}$ are universal scaling functions, {``$\perp$" denotes directions perpendicular to $\hat{\bf x}$,} $\gamma$ is a model-dependent non-universal speed, and the universal scaling exponents are given by
\begin{subequations}
\label{eq:allexponents}
\begin{align}
&\zeta={d+1\over 3}={4\over3} \ , \ \ \chi={2-d\over 3}=-{1\over3}\,,
\\
&\zeta'={2(d+1)\over d+7}={4\over5}\,,\ \
z'={4(d+1)\over d+7}={8\over5}\,, 
\\
&\chi'=-\left({d^2+4d-9\over 2(d+7)}\right)=-{3\over5}\,,
\end{align}
\end{subequations}
 for spatial dimensions between $2$ and $5$,
where the final equalities hold in the physically relevant case
$d=3$.

{\it Hydrodynamic description.---}We start with the hydrodynamic equation of motion (EOM) of a generic incompressible polar active fluid  with  both quenched and annealed fluctuations constructed using symmetry arguments \cite{toner_prl95,toner_pre98}. The \emph{only} hydrodynamic variable we need to account for is the velocity field ${\bf v}$. However,  in contrast to the  Navier-Stokes equations for passive incompressible fluids, ${\bf v}$ is hydrodynamic not because it is conserved -- it is not, since momentum is not conserved -- but because it is a broken symmetry variable (more precisely certain components of it are). Our EOM also contains terms that violate
momentum conservation and Galilean invariance,
because the motile agents move through a frictional (and  disordered) medium.
Furthermore, because the system is non-equilibrium, many terms forbidden in equilibrium are allowed here as well \cite{LPDJSTAT}.
These  considerations imply the following EOM  \cite{toner_prl95,toner_pre98}:
\beqn
\label{eq:maineom}
\pp_t \bv+ \lambda_1 (\bv \cdot \vnab )\bv &=& -\vnab P  -(\bv \cdot \vnab P_1) \bv +U(|\bv|) \bv
\\
\nonumber
&&+\mu_1 \nabla^2 \bv +\mu_2 (\bv \cdot \vnab)^2 \bv
 +\bff_{_Q}+\bff_{_A}
\ ,
\eeqn
where the ``pressure" $P$ acts as a Lagrange multiplier to enforce the incompressibility constraint: $\vnab \cdot \bv = 0$, the ``anisotropic" pressure is an arbitrary function of the speed $|\bv|$, and $U(|\bv|)<0$ for $|\bv|>v_0$ and $U(|\bv|)>0$ for $|\bv|<v_0$; these last two inequalities ensure that the system has a non-zero preferred speed $v_0$, which allows it to be in the ordered phase. Furthermore, $\bff_{_Q}$ and $\bff_{_A}$ are respectively the quenched and annealed noises, which have  zero means and correlations of the form
\begin{subequations}
\begin{align}
\langle f_{_Q}^i(\br,t)f_{_Q}^j(\br',t')\rangle&=
2D_{_Q}\delta_{ij}\delta^d(\br-\br')\,,\\
\langle f_{_A}^i(\br,t)f_{_A}^j(\br',t')\rangle&=
2D_{_A}\delta_{ij}\delta^d(\br-\br')\delta(t-t')\, ,
\end{align}
\end{subequations}
where the indices $i,j$ enumerate the spatial coordinates. In the EOM (\ref{eq:maineom}), we have only included terms that are relevant to the universal scaling behavior,  based on the DRG analysis below.

{We focus on} the broken-symmetry moving phase, {and} consider the {local} velocity deviation $\bu(\br,t)$, from the mean flow $v_0 \hat{\bf x}$: $\bu = \bv- v_0\hat{\bf x}$,
{whose EOM is obtained from (\ref{eq:maineom}) by}
keeping only relevant terms (some of which, however, are non-linear):
\begin{subequations}
\label{eq:lin_u}
\begin{align}
\pp_t u_x=&-\pp_xP-(\gamma+b)\pp_xu_x-\alpha\left(u_x+{u^2\over 2v_0}\right)+f_{_Q}^x+f_{_A}^x\,,\label{Genx}\\
\pp_t \bu_{_\perp}=&-\vnab_{_\perp} P-\gamma\pp_x\bu_{_\perp}-\lambda_1(\bu_{_\perp} \cdot \vnab_{_\perp})\bu_{_\perp}+\bff_{_Q}^\perp+\bff_{_A}^\perp
\nonumber
\\
&-{\alpha\over v_0}\left(u_x+{u^2\over 2v_0}\right)\bu_{_\perp}
+\mu_{_\perp}\nabla^2_{_\perp}\bu_{_\perp}+\mu_x\pp_x^2\bu_{_\perp}\,.
\label{Genperp}
\end{align}
\end{subequations}
where $\gamma\equiv\lambda_1v_0$,
$\alpha\equiv -v_0\left(\dd U\over\dd |\bv|\right)_{|\bv|=v_0}$,
$b\equiv v_0^2\left(\dd P_1\over\dd |\bv|\right)_{|\bv|=v_0}$,
$\mu_{_\perp}=\mu_1$, and $\mu_x=\mu_1+\mu_2v_0^2$.

{\it Linear theory.---}First we examine the linearized version of (\ref{Genx}) and (\ref{Genperp}). {In terms of} the spatiotemporally Fourier transformed field  $\bu(\bq, \omega) = (2\pi)^{-(d+1)/2} \int \dd t \dd^d r\,  \ee^{{\ctea -}\ii(\bq \cdot \bbr-\omega t)}\bu(\bbr, t)$,  the linearized EOMs read
\begin{subequations}
\begin{align}
\Big[-\ii(\omega-(\gamma +b)q_x) +\alpha\Big] u_x&=-\ii q_xP+f_{_Q}^x+f_{_A}^x\,,\label{linearx}\\
\Big[-\ii(\omega-\gamma q_x) +\Gamma(\bq) \Big] u_{_L}&=-\ii q_{_\perp} P+f_{_Q}^L+f_{_A}^L\,,
\label{linearl}\\
\Big[-\ii(\omega-\gamma q_x) +\Gamma(\bq) \Big] \bu_{_T} &= \bff_{_Q}^T+\bff_{_A}^T\, ,
\label{lineart}
\end{align}
\end{subequations}
where we have
decomposed $\bu_{_\perp}$ as
$\bu_{_\perp}(\bq, \omega)=u_{_L}(\bq, \omega)\hat{\bq}_{_\perp}+\bu_{_T}(\bq, \omega)$, and introduced the $\bq$-dependent damping coefficient:
\beq
\Gamma(\bq)\equiv \mu_{_\perp} q^2_{_\perp}+\mu_xq_x^2\ .
\eeq
We now calculate the autocorrelation functions in this linear theory. Since the EOM of $\bu_{_T}$ is completely decoupled from the other two modes, its autocorrelation function
can be obtained immediately
\beqn
\langle \bu_{_T}(\bq,\omega)\cdot\bu_{_T}(\bq', \omega')\rangle
&=&C_{_{A}}^T(\bq,\omega)\delta(\omega+\omega')\delta(\bq+\bq')\nonumber\\
&+&C_{_{Q}}^T(\bq)\delta(\omega)\delta(\omega')\delta(\bq+\bq')\, ,
\label{Linear_tr1}
\eeqn
where
\begin{subequations}
\label{Linear_tr2}
\begin{align}
C_{_{A}}^T(\bq,\omega) &= {2D_{_A}(d-2)\over \left(\omega-\gamma q_x\right)^2+\left[\Gamma(\bq)\right]^2}\,,\\
C_{_{Q}}^T(\bq)&= {4\pi(d-2) D_{_Q}\over {\gamma^2q_x^2+\left[\Gamma(\bq)\right]^2}}\,,
\end{align}
\end{subequations}
and the subscripts $A$ and $Q$ denote the annealed and quenched parts, respectively. The correlation of $\bu_{_T}$ constitutes the most divergent part of the velocity correlator, since $u_x$ is the ``massive" mode, and $u_{_L}$  is ``almost massive" because  $u_x$ is enslaved to it by the incompressibility condition $q_xu_x+q_{_\perp} u_{_L}=0$. This renders it impossible, for most directions of $\bq$, to create a non-zero $u_{_L}$ without also creating a massive $u_x$ field along with it. We explicitly calculate the autocorrelations of $u_x$ and $u_{_L}$ in \cite{SM}.

Using (\ref{Linear_tr1}) and (\ref{Linear_tr2}),
the fluctuations of $\bu$ in real space
and time can be obtained by integrating over all wavevectors $\bq$ and frequencies $\omega$. Performing the frequency integral gives
\beq
\langle |\bu(\br, t)|^2\rangle
= {(d-2)\over (2\pi)^d}\int \dd^d \bq
\left[{D_{_A}\over \Gamma(\bq)}
+{2 D_{_Q}\over {\gamma^2q_x^2+\left[\Gamma(\bq)\right]^2}}\right]\, . \label{3Dlinear_Real_uu}
\eeq
In the infrared limit ($\bq \rightarrow {\bf 0}$), the second term in the integrand  (due to the quenched disorder $D_{_Q}$) is more divergent and thus dominates the fluctuations in the system. The integral {of this term is} logarithmically divergent in $d=3$, {which implies} quasi-long-range orientational order at this lower critical dimension.
Further,  the scaling of (\ref{Linear_tr2}) and (\ref{3Dlinear_Real_uu}) yields   the scaling exponents {for the quenched and annealed fluctuations} in this linear theory:
\begin{subequations}
\label{Linear_Exp}
\begin{align}
\zeta_{\rm lin}=2,~~~\chi_{\rm lin}={3-d\over 2}\,, \label{Linear_Exp_Q}\\
{\zeta'_{\rm lin}=1,~~~\chi'_{\rm lin}={2-d\over 2},~~~z'_{\rm lin}=2}\,. \label{Linear_Exp_A}
\end{align}
\end{subequations}
However, all of the above conclusions are modified by the nonlinearity in the EOM when $d<5$. In particular, the flock moves coherently, i.e., has long-range order, for all $d>2$. That is, non-linearities change the lower critical dimension $d_{_{LC}}$ of this system from the linear theory's prediction $d_{_{LC}}=3$ to $d_{_{LC}}=2$.

{\it Nonlinear theory.---}As indicated by the linear theory, fluctuations in $\bu$ are dominated by those of $\bu_{_\perp}$ (more precisely the transverse components of $\bu_{_\perp}$, i.e., ${\bf u}_{_T}$).
{The full EOM of  $\bu_{_\perp}$ (\ref{Genperp}), after eliminating all irrelevant terms \cite{SM}, becomes}
\beqn
\nonumber
\pp_t \bu_{_\perp}&=&-\vnab_{_\perp} P-\gamma\pp_x\bu_{_\perp}-\lambda_1(\bu_{_\perp} \cdot \vnab_{_\perp})\bu_{_\perp}+\mu_{_\perp}\nabla_{_\perp}^2\bu_{_\perp}
\\
&&+\mu_x\pp_x^2\bu_{_\perp}+
\bff_{_Q}^\perp+\bff_{_A}^\perp\,.
\label{Genperp3D3}
\eeqn
We   {will} now obtain exact scaling exponents from (\ref{Genperp3D3}) using a DRG argument \cite{forster_pra77}.
In this DRG analysis, we first decompose the field $\bu_{_\perp}$ into the rapidly varying and slowly varying parts, which are supported in the small- and large-momentum space respectively. We then average the EOM over the rapidly varying fields to get an effective EOM for the slowly varying fields. In this process the various coefficients in the EOM get renormalized and this renormalization can be represented by Feynman diagrams. We will therefore refer to
all corrections that arise due to this part of the DRG process as ``graphical corrections".
Next we rescale the time, lengths, and the field as follows:
\beq
t\to te^{z\ell},~~x\to xe^{\zeta\ell},~~r_{_\perp}\to r_{_\perp} e^{\ell},~~\bu_{_\perp}\to \bu_{_\perp} e^{\chi\ell}
\ ,
\label{Rescale1}
\eeq
to restore the supporting momentum space (i.e., the Brillouin zone) back to its original  size.
This procedure is repeated infinitely, leading to the following recursion relations for the various coefficients:
\begin{subequations}
\label{eq:drg}
\begin{align}
{\dd\mu_{_\perp}\over\dd\ell} &=\left(z-2+\eta_{_\perp}\right)\mu_{_\perp}\,,\label{3Dflow _mu_perp}\\
{\dd\mu_x\over\dd\ell}&=\left(z-2\zeta\right)\mu_x\,,\label{3Dflow_mu_x}\\
{\dd\gamma\over\dd\ell}&=\left(z-\zeta\right)\gamma\,,\label{3Dflow_gamma}\\
{\dd\lambda_1\over\dd\ell}&=\left(z+\chi-1\right)\lambda_1\label{3Dflow_lambda_1}\,,\\
{\dd D_{_Q}\over\dd\ell}&=\left[2z-2\chi-\zeta-(d-1)\right]D_{_Q}\,,\label{3Dflow_D_Q}\\
{\dd D_{_A}\over\dd\ell}&=\left[z-2\chi-\zeta-(d-1)\right]D_{_A}\,.\label{3Dflow_D_A}
\end{align}
\end{subequations}
where $\eta_{_\perp}$ represents the graphical correction to $\mu_{_\perp}$ -- the only graphical correction to the DRG flow equations above. We explain why there are no other graphical corrections {in the SM \cite{SM}}.

Having established the form of the DRG recursion relations (\ref{eq:drg}), - that is, the fact that {\it only} $\mu_\perp$ gets any graphical corrections (those denoted by $\eta_\perp$ in (\ref{3Dflow _mu_perp})) -
we will now show that the quenched disorder is {\it always} relevant at the ``annealed" fixed point that controls the ordered phase in the absence of quenched disorder,  even when graphical corrections are taken into account, and  determine the universal scaling exponents (\ref{eq:allexponents}) in the presence of quenched disorder {\it exactly}.

Note that the   {\it form} of the recursion relations is exactly the same in the absence of quenched disorder as in its presence; that is, the recursion relations (\ref{eq:drg}) continue to hold,   albeit with different values for $\eta_{_\perp}$ depending on whether quenched disorder is present or not. This is because the arguments presented in the SM \cite{SM} for the quenched problem apply equally well to the annealed problem.   (The argument for the non-renormalization of $\lambda_1$ is different in the annealed case \cite{us}, but the result stands.)
Therefore,   the same conclusion holds:
only $\mu_{_\perp}$ gets graphically corrected. The only differences that the absence of quenched disorder makes are: 1)  the graphical correction $\eta_{_\perp}$ will now be generated entirely by the {\it annealed} noise, rather than the quenched noise, and 2) the values of the exponents $z$, $\zeta$, and $\chi$ will change to the values found in the study of the annealed problem \cite{chen_njp18}). They did so by
 choosing $z$, $\zeta$, and $\chi$ to fix $\mu_x$, $\mu_{_\perp}$, and $D_{_A}$, since those parameters control the dominant fluctuations in the absence of quenched disorder.  To see that only these parameters matter in the annealed problem, one need simply inspect the annealed contribution (i.e., the $D_{_A}$ term) in (\ref{3Dlinear_Real_uu}).

Making this choice, and noting that the DRG eigenvalues of $D_{_A}$ and $D_{_Q}$ (i.e., the terms in square brackets in equations (\ref{3Dflow_D_Q}) and (\ref{3Dflow_D_A})) differ by precisely $z$, it follows that, since we are choosing
 $z$, $\zeta$, and $\chi$ to make the DRG eigenvalue of   $D_{_A}$
vanish, that the eigenvalue for $D_{_Q}$ is given by $z$. Since $z$ is always positive ($z={2(d+1)\over5}$ for $d\le4$ and $z=2$ for $d>4$ \cite{chen_njp18}), it follows that the quenched noise is always strongly relevant{; i.e., it} will change the long-distance and time scaling of fluctuations.

We can calculate the new scaling that ensues in the presence of {\it quenched} noise by much the same reasoning that we just outlined for the annealed problem. The only change is that it is now  $\gamma$, $\mu_{_\perp}$, and $D_{_Q}$ that we must keep constant at this fixed point, since they control the dominant (i.e., quenched) fluctuations  in (\ref{3Dlinear_Real_uu}). The coefficient of the relevant non-linear term $\lambda_1$ must also be fixed at this stable fixed point. This reasoning  leads to four linear equations:
\begin{subequations}
\begin{align}
z-2+\eta_{_\perp}=0 \ &, \ \ \
z-\zeta=0\,,\\
2z-2\chi-\zeta-(d-1)=0\, &,\ \ \
z+\chi-1=0\,.
\end{align}
\end{subequations}
Solving these equations we find
\beq
z=\zeta={d+1\over 3}\,,~~~~~~~~\chi={2-d\over 3}\,,
~~~~~~~~\eta_{_\perp}={5-d\over 3}\,.\label{Exponents3D1}
\eeq
We see that $\zeta$ and $\chi$ differ from those obtained from the linear theory (\ref{Linear_Exp_Q}),  and only become equal to those linear values at the upper critical dimension $d=5$. Furthermore, $\chi<0$ {which implies} long-range order, for all $d>2$. At exactly two dimensions, our present analysis no longer holds since the only ``soft" dimension  is coupled directly to the ``hard" dimension (i.e., along the direction of collection motion)
through the incompressibility condition, and a completely different formulation of the problem is required, as described in \cite{us}. We note that  $d=2$ is also a singular limit of  incompressible flocks {\it without} quenched disorder \cite{chen_natcomm16,chen_njp18} (see Fig.~1of the SM \cite{SM}).

{\it Scaling behavior.---}Using the exponents {(\ref{Exponents3D1})} we now derive {the $\bu$-$\bu$ correlation function
(\ref{3Dcorrelation1}) and the exponents (\ref{eq:allexponents})}, and discuss the scaling behavior of the correlation function in different limits.
The dominant part of  the $\bu$-$\bu$ correlation function {in Fourier space} is displayed in (\ref{Linear_tr1}), with $\mu_x$, $\gamma$, and $D_{_{A,Q}}$
{given by their ``bare" values, since there are no graphical corrections to them,} $\mu_{_\perp}$
{is now a} $\bq$-dependent
quantity:
\beqn
\mu_{_\perp}(\bq)=\mu_{_{\perp 0}}\left(q_{_\perp}\over\Lambda\right)^{-\eta_{_\perp}}f_{\mu_{_\perp}}\left(q_x/\Lambda'\over \left(q_{_\perp}/\Lambda\right)^{\zeta}\right)\,,\label{Anomalous_mu_perp}
\eeqn
where $f_{\mu_{_\perp}}$ is a scaling function such that
\beq f_{\mu_{_\perp}}(s)
\propto\left\{
\begin{array}{ll}
{\rm constant}\,,&s \ll 1\,,\\
s^{-\eta_{_\perp}/\zeta}\,,&s \gg 1\,.
\end{array}
\right.
\eeq
{Here}
$\Lambda$ is the non-universal ultra-violet cutoff, and $\Lambda'={\mu_{_{\perp0}}\over\gamma}\Lambda^2$. The subscript ``0" in $\mu_{\perp}$ denotes the bare value.

{Fourier-transforming $\langle \bu_{_T}(\bq,\omega)\cdot\bu_{_T}(\bq', \omega')\rangle$, we obtain}
\beq
\langle \bu(\br, t)\cdot\bu(\mathbf 0, 0)\rangle
=C_{_{A}}(\br, t)+C_{_{Q}}(\br)\,,\label{uu1}
\eeq
where
\begin{subequations}
\begin{align}
\nonumber
C_{_{A}}(\br, t)
=&\int \frac{\dd\omega\dd^d q}{(2\pi)^{d+1}}
\ee^{\ii(\bq\cdot\br-\omega t)}
\\
& \times
\left\{{2(d-2)D_{_A}\over \left(\omega-\gamma q_x\right)^2+\left[\mu_xq_x^2+
\mu_{_\perp}(\bq)q_{_\perp}^2\right]^2}\right\}
\,,
\label{C3A1}
\\
C_{_{Q}}(\br)
=&\int \frac{\dd^d q}{(2\pi)^d}
\left\{{2(d-2) D_{_Q}\over {\gamma^2q_x^2+\left[\mu_{_\perp}(\bq)q_{_\perp}^2\right]^2}}\right\}\ee^{\ii\bq\cdot\br}\, .
\label{Nonlinear_real_uu_Quench}
\end{align}
\end{subequations}
are the correlations coming from the annealed and quenched noises, respectively.

For $C_{_{Q}}(\br, t)$, by changing the variables of integration
{to: ${\bk}_{_\perp} \equiv \bq_{_\perp} (r_{_\perp}\Lambda)$ and
$k_x \equiv q_{_x} (r_{_\perp}\Lambda)^{\zeta}$},
(\ref{Nonlinear_real_uu_Quench}) can be written as
\beqn
C_{_{Q}}(\br)=r_{_\perp}^{2\chi}G_{_{Q}}\left(|x|\over r_{_\perp}^{\zeta}\right)\,.\label{C_Q1}
\eeqn
 where $G_{_Q}$ is a  scaling function {given} in \cite{SM}.

For $C_{_{A}}(\br, t)$,  the annealed part of the correlation function, the dominant contribution to the integral in (\ref{C3A1}) comes from the region in which the two terms inside the square brackets in the denominator become comparable:
\beqn
\mu_{x0}q_x^2 \sim \mu_{_\perp}(\bq)q_{_\perp}^2\,.\label{Dregion}
\eeqn
Since $\mu_{_\perp}(\bq)$ diverges at small $\bq$ [see (\ref{Anomalous_mu_perp})], (\ref{Dregion}) implies $q_x\gg q_{_\perp}$ and hence $q_x\gg q_{_\perp}^\zeta$ since $\zeta>1$ for $d>2$ [see (\ref{Exponents3D1})]. Using this in (\ref{Anomalous_mu_perp}) we get
\beqn
\mu_{_\perp}(\bq)=\mu_{_{\perp 0}}\left(q_{_x}\over\Lambda'\right)^{-{\eta_{_\perp}\over\zeta}}\,.\label{Anomalous_mu_perp1}
\eeqn
Inserting (\ref{Anomalous_mu_perp1}) into (\ref{C3A1}) , {introducing $\omega'=\omega-\gamma q_x$, and further changing variables of integration{: $\bk_{_\perp} \equiv \bq_{_\perp} r_{_\perp}$, $k_x \equiv q_{_x} (r_{_\perp}\Lambda)^{\zeta'}$, $\Omega \equiv \omega' (r_{_\perp}\Lambda)^{z'}$,}}
we obtain
\beqn
C_{_{A}}(\br, t)=r_{_\perp}^{2\chi'}G_{_{A}}\left({|x-\gamma_0t|\over r_{_\perp}^{\zeta'}},{|t|\over r_{_\perp}^{z'}}\right)\,,\label{C_A1}
\eeqn
where $\zeta'$, $z'$, $\chi'$ are given in (\ref{eq:allexponents}b), {and $G_{_A}$ is a  scaling function given in \cite{SM}.}

Inserting (\ref{C_Q1}) and (\ref{C_A1}) into (\ref{uu1}) gives (\ref{3Dcorrelation1}). We now delineate its scaling behaviour in distinct regimes.
Since $\chi>\chi'$ and ${\chi\over\zeta}>{\chi'\over\zeta'}$, the equal-time correlation is dominated by the contribution from the quenched fluctuations. Specifically,
\beqn
\nonumber
\langle \bu(\br,0)\cdot\bu(\mathbf 0,0)\rangle&=&r_{_\perp}^{2\chi}G_{_Q}\left(x\over r_{_\perp}^{\zeta}\right)
\\
&\propto&\left\{
\begin{array}{ll}
r_{_\perp}^{2\chi}\,,&|x|\ll r_{_\perp}^\zeta\,,\\
|x|^{2\chi\over\zeta}\,,&|x|\gg r_{_\perp}^\zeta\,.
\end{array}
\right.
\eeqn
On the other hand, the time-dependence of the correlation is solely determined by the  annealed fluctuations, since the quenched fluctuations are constant in time. However, the quenched fluctuations do affect the equal-position correlation indirectly by renormalizing the diffusion coefficient $\mu_{_\perp}$, which is one of the controlling parameters of the annealed fluctuations (see (\ref{C3A1})). {As a result, the difference between the equal positon correlation function at time $t$ and its value at $t=0$ is given by}
\beqn
\langle \bu(\mathbf 0, t)\cdot\bu(\mathbf 0, 0)\rangle-{\langle \bu(\mathbf 0, 0)\cdot\bu(\mathbf 0, 0)\rangle}=C_{_{A}}(\mathbf 0, t)=A|t|^{\theta}\,,\nonumber\\
\label{eq:theta}
\eeqn
where $A$ is a non-universal constant and
\beqn
\theta ={2\chi'\over \zeta' }=-\left[{d^2+4d-9\over 2(d+1)}\right]= -{3\over 2}\,,
\eeqn
with the last equality holding in the physical case $d=3$. {We give the detailed argument for this expression for $\theta$ in the SM \cite{SM}.}
In Fig.~1 of the SM \cite{SM}, we show how some of the scaling exponents vary with spatial dimension, and how they compare with those in the purely annealed case \cite{chen_natcomm16, chen_njp18}.

{\it Summary \& Outlook.---}We have considered the effects of quenched disorder in incompressible polar active fluids in the flocking phase, and showed that quenched  disorder
 make  the scaling behavior of the system  very different from that predicted by linearized hydrodynamics, and from that of an incompressible polar active fluid with only annealed disorder.
While this work focuses on an one-component active fluid in the incompressible limit,
an  interesting future direction would be to consider the hydrodynamic behavior of active suspensions,
which are   two-component (swimmers and solvent) systems that are only incompressible as a whole.

\begin{acknowledgments}
    {\it Acknowledgements.---}J.T.
 thanks the Max Planck Institute for the Physics of Complex Systems,
Dresden, Germany, for their support through a Martin
Gutzwiller Fellowship during the period this work was underway. He also thanks the Coll\`{e}ge de France for their hospitality during a visit that lead to the start of this work. L.C. acknowledges support by the National Science
Foundation of China (under Grant No. 11874420), and  thanks the MPI-PKS, where the early stages of this work were done, for their support. We all thank Wanming Qi for calling our attention to the lack of pseudo-Galilean invariance in the presence of quenched disorder. AM was supported by a TALENT fellowship awarded by the CY Cergy Paris universit\'e.
\end{acknowledgments}

\onecolumngrid
\appendix
%
\section{Autocorrelation functions for $u_x$ and $u_{_L}$}
In this supplementary section, we display the autocorrelation functions for $u_x$ and $u_L$ obtained from Eq.~(6) of the main text.
For this, we eliminate the pressure $P$ via the incompressibility condition $q_x u_x+q_\perp u_{_L}=0$ and find that {the} autocorrelation functions, $\langle u_m(\bq,\omega)u_m(\bq', \omega')\rangle$  where $m=x, L$, are of the form
\begin{align}
	C^m_{_{A}}(\bq,\omega)\delta(\omega+\omega')\delta(\bq+\bq') +
	C^m_{_{Q}}(\bq)\delta(\omega)\delta(\omega')\delta(\bq+\bq')
	\ ,
\end{align}
{where}
\begin{subequations}
	\label{eq:Ca&Cq}
	\begin{align}
		C_{_{A}}^{x,L}(\bq,\omega)&={q_{\perp,x}^2\over q^2}{2D_{_A}\over \left[\omega-\left({b q_{_\perp}^2\over q^2}+\gamma\right)q_x\right]^2+\left[{\alpha q_{_\perp}^2\over q^2}+{q_x^2\over q^2}\Gamma(\bq)\right]^2}\,,\\
		C_{_{Q}}^{x,L}(\bq)&={q_{\perp,x}^2\over q^2}{4\pi D_{_Q}\over \left({b q_{_\perp}^2\over q^2}+\gamma\right)^2q_x^2+\left[{\alpha q_{_\perp}^2\over q^2}+{q_x^2\over q^2}\Gamma(\bq)\right]^2}\, .
	\end{align}
\end{subequations}
In the hydrodynamic limit ($\omega\to 0$, $\bq\to\mathbf 0$), we see that $C_{_{A,Q}}^{x,L}$ are finite for most $\bq$'s (i.e., when $q_x {\sim}\ q_{_\perp}$), while, as shown in the main text, $C_{_{A,Q}}^T$ diverges as either $1/\omega^2$ or $1/q^2$. Therefore, as we argued in the main text, $\bu$-$\bu$ correlation is dominated by the fluctuations in $\bu_T$,  i.e.,
\beq
\langle \bu(\bq,\omega)\cdot\bu(\bq', \omega')\rangle
\approx\langle \bu_{_T}(\bq,\omega)\cdot\bu_{_T}(\bq', \omega')\rangle
\ .\label{Linear_Fourier_uu}
\eeq

\section{Detailed argument for non-renormalisation of $\lambda_1$}
In {this} {section}, we present our argument that $\lambda_1$ gets no graphical corrections.
Consider first, by way of illustration, the {\it  homogeneous} Navier-Stokes equation (i.e., one with no noise or other external force), but with a parameter $\lambda_1 \ne 1$ introduced (say, to allow us to do rescaling in the DRG \cite{FNS}). ``Galilean invariance" means that if we have a solution $\bv^{(1)}(\br,t)$ of the Navier-Stokes equation, which now reads

\beq
\pp_t \bv + \lambda_1 \bv \cdot\nabla\bv-\nu \nabla^2 \bv={\bf 0} \,,
\label{NS}
\eeq
then there's a family of other solutions parameterized by a vector $\bw$ given by
\beq
\bv^{(\bw)}(\br,t)=\bv^{(1)}(\br-\lambda_1 \bw t , t)+\bw
\label{boost}
\eeq
which also satisfies the same equation, with the same value of $\lambda_1$.

But now, if we consider the {\it in}homogeneous Navier-Stokes equation
\beq
\label{inhomNS}
\pp_t \bv(\br,t) + \lambda_1 \bv(\br,t) \cdot\nabla \bv(\br,t)-\nu \nabla^2 \bv(\br,t)=\bff(\br,t),
\eeq
and have a solution $\bv^{(1)}(\br,t)$ to this equation, then
\beq\label{Galtrans}
\bv^{(\bw)}(\br,t)=\bv^{(1)}(\br-\lambda_1 \bw t , t)+\bw
\eeq
is not a solution of \eqref{inhomNS}. Indeed, if we plug \eqref{Galtrans} into the left hand side of \eqref{inhomNS}, we get
\bew
\beq
\pp_t \bv^{(1)}(\br-\lambda_1 \bw t , t) + \lambda_1 \bv^{(1)}(\br-\lambda_1 \bw t , t) \cdot\nabla \bv^{(1)}(\br-\lambda_1 \bw t , t)-\nu \nabla^2 \bv^{(1)}(\br-\lambda_1 \bw t , t)
\eeq
\ew
which, according to equation \eqref{inhomNS}, equals $\bff(\br-\lambda_1 \bw t, t)$. Thus, plugging our trial solution \eqref{Galtrans} into \eqref{inhomNS} leads to
\beq
\bff(\br-\lambda_1 \bw t, t)=\bff(\br,t)
\eeq
which is clearly not true in general.

To say this another way, if ${\bf v}^{(1)}$ obeys \eqref{NS}, then $\bv^{(\bw)}(\br,t)$ obeys
\bew
\beq
\label{inhomtrans}
\pp_t \bv^{(1)}(\br-\lambda_1 \bw t , t) + \lambda_1 \bv^{(1)}(\br-\lambda_1 \bw t , t) \cdot\nabla \bv^{(1)}(\br-\lambda_1 \bw t , t)-\nu \nabla^2 \bv^{(1)}(\br-\lambda_1 \bw t , t)=\bff(\br-\lambda_1 \bw t, t)
\eeq
\ew
which is a different equation, because it has a different force on the right hand side.

However, Galilean invariance {\it does} work in a statistical sense: if we look at the correlations of the noise in \eqref{inhomtrans} (i.e., $\bff(\br-\lambda_1 \bw t, t))$, and assume that $\bff(\br,t)$ satisfies local white noise (annealed) statistics
\beq\label{Galnoiseorig}
\la f_i(\br,t) f_j(\br,t')\ra=D \delta_{ij} \delta^d(\br-\br') \delta(t-t')
\eeq
then
\bew
\beq
\label{Galnoise}
\la  f_i(\br-\lambda_1 \bw t, t) f_j(\br'-\lambda_1 \bw t', t')\ra=D \delta_{ij} \delta[\br-\br'-\lambda_1 \bw (t-t')] \delta(t-t')
\eeq
\ew
However, since the $\delta(t-t')$ in this expression is only non-zero when $t=t'$, it follows that we can set $t-t'=0$ in the $\delta[\br-\br'-\lambda_1 \bw (t-t')]$ without changing anything (since doing so only changes the value of that delta function when $t \ne t' $, in which case $\delta[\br-\br'-\lambda_1 \bw (t-t')]$ is multiplied by zero anyway.
So, setting $t-t'=0$ in the $\delta[\br-\br'-\lambda_1 \bw (t-t')]$, Eq.~\eqref{Galnoise} becomes
\beq
\la f_i(\br-\lambda_1 \bw t, t) f_j(\br'-\lambda_1 \bw t', t')\ra=D \delta_{ij} \delta(\br-\br') \delta(t-t')
\eeq
which has the same statistics \eqref{Galnoiseorig} as the noise in the original equation. So the inhomogeneous Navier-Stokes equation with annealed noise is Galilean invariant in a statistical sense.

However, this argument does not work if the noise is quenched. In this case, going through the above manipulations leads to the replacement of the original quenched noise statistics
\beq\label{Qnoise0}
\la f_i(\br,t) f_j(\br',t')\ra=D \delta_{ij} \delta^d(\br-\br')
\eeq
with
\beq\label{GalnoiseQ}
\la f_i(\br-\lambda_1 \bw t, t) f_j(\br'-\lambda_1 \bw t', t')\ra=D \delta_{ij} \delta[\br-\br'-\lambda_1 \bw (t-t')]
\eeq
and now there's no convenient second delta function $\delta(t-t') $ to allow us to set $t=t'$ in $\delta[\br-\br'-\lambda_1 \bw (t-t')]$. Therefore, the statistics \eqref{GalnoiseQ} really are different from the statistics \eqref{Qnoise0}, so we can not argue for even statistical Galilean invariance.

However, despite this lack of Galilean invariance (and consequently, pseudo-Galilean invariance \cite{NL}), $\lambda_1$ does not {\ctea \sout{get}} receive any graphical correction even in the presence of quenched noise.

To see this, let us consider the steady-state equation of motion, which is sufficient since the static quenched noise dominates. This looks very much like the Navier-Stokes equation, with $x$ playing the role of time, and $r_\perp$ playing the role of space. To be precise, it reads
\bew
\beq
\gamma \pp_x \bu_\perp(\br_\perp, x) +\lambda_1 (\bu_\perp(\br_\perp, x) \cdot\nabla_\perp) \bu_\perp(\br_\perp, x) - \mu_\perp \nabla_\perp^2 \bu_\perp(\br_\perp, x)=\bff_Q(\br_\perp, x)
\label{quenchedEOM}
\eeq
\ew
with $\bff_{_Q}$ having statistics
\beq\label{Qnoise}
\la f_{_Q}^i(\br_\perp,x) f_{_Q}^j(\br_\perp,x') \ra=D_{_Q} \delta_{ij} \delta(\br_\perp-\br'_\perp) \delta(x-x')\,.
\eeq
In writing equation (\ref{quenchedEOM}), we have neglected the term $\mu_x \pp_x^2 \bu_\perp(\br_\perp, x)$ that we retain in the main text, because that term proves to be irrelevant for the quenched fluctuations that dominate the graphical corrections.
Now assume we have a solution $\bu_\perp^{(1)}(\br_\perp, x)$ of \eqref{quenchedEOM}.
Then the ``very-pseudo Galilean boosted" vector function obtained by treating $x$ the way we treated time in the Navier-Stokes equation, namely
\beq
\label{verypseudo}
\bu_\perp^{(\bw)}(\br,t)=\bu_\perp^{(1)}(\br_\perp-(\lambda_1/\gamma) \bw x , x)+\bw
\eeq
satisfies
\beq
\gamma \pp_x \bu_\perp(\br_\perp, x) +\lambda_1 (\bu_\perp(\br_\perp, x) \cdot\nabla_\perp) \bu_\perp(\br_\perp, x) - \mu_\perp \nabla_\perp^2 \bu_\perp(\br_\perp, x) - \mu_x \pp_x^2 \bu_\perp(\br_\perp, x) =\bff_Q(\br_\perp-(\lambda_1/\gamma) \bw x, x)
\ .
\eeq
The statistics of the noise on the right hand side are the same as as those of the original noise, by an argument almost identical to the one used above for the annealed Navier-Stokes equation. Specifically, we have
\bew
\beq
\label{PseudoGalnoi}
\la f_{_Q}^i(\br_\perp-(\lambda_1/\gamma) \bw x, x) f_{_Q}^j(\br'_\perp-(\lambda_1/\gamma) \bw x', x')\ra=D_{_Q} \delta_{ij} \delta[\br_\perp-\br'_\perp-(\lambda_1/\gamma) \bw (x-x')] \delta(x-x')
\eeq
\ew
As in the Navier-Stokes case, here we can replace $x-x'$ with zero in $\delta[\br_\perp-\br'_\perp-(\lambda_1/\gamma) \bw (x-x')]$,  since $\delta(x-x')$ will vanish when
$x \ne x'$.

Making that replacement, \eqref{PseudoGalnoi} becomes
\bew
\beq
\la f_{_Q}^i(\br_\perp-(\lambda_1/\gamma) \bw x, x) f_{_Q}^j(\br'_\perp-(\lambda_1/\gamma) \bw x', x')\ra=D_{_Q} \delta_{ij} \delta(\br_\perp-\br'_\perp) \delta(x-x')
\eeq
\ew
which are exactly the same statistics \eqref{Qnoise} as the original noise. Hence, the static equation of motion has the ``very-pseudo-Galilean invariance" \eqref{verypseudo} in a statistical sense. Since this invariance involves $\lambda_1/\gamma$, it follows that $\lambda_1/\gamma$ can not get any graphical corrections.

{We now note that the relevant part of the nonlinear term $(\bu_{_\perp} \cdot \vnab_{_\perp})\bu_{_\perp}$ is a total ``$\perp$" derivative:
\beqn
(\bu_{_\perp} \cdot \vnab_{_\perp})\bu_{_\perp}
&=&\pp_i^\perp\left(u_i^\perp\bu_{_\perp}\right)-\left(\nabla_{_\perp}\cdot\bu_{_\perp}\right)\bu_{_\perp}
\\
\nonumber
&=&\pp_i^\perp\left(u_i^\perp\bu_{_\perp}\right)+\left(\pp_x u_x\right)\bu_{_\perp}
\approx \pp_i^\perp\left(u_i^\perp\bu_{_\perp}\right)\,,
\eeqn
where in the second equality we have used the incompressibility constraint $\nabla_{_\perp}\cdot u_{_\perp}+\pp_xu_x=0$, and in ``$\approx$" we have neglected $\left(\pp_x u_x\right)\bu_{_\perp}$ since it is irrelevant compared to $\pp_i^\perp\left(u_i^\perp\bu_{_\perp}\right)$.

Since this sole relevant non-linearity is a total $\perp$-derivative, there are no graphical corrections to $\mu_2$, $\gamma$, and $D_{_{A,Q}}$,  because Feynman diagrams constructed from $\pp_i^\perp\left(u_i^\perp\bu_{_\perp}\right)$ can only generate corrections to terms that are themselves  $\perp$ derivatives of some quantity. Since none of the terms $\pp_x^2\bu_{_\perp}$, $\pp_x\bu_{_\perp}$, and the noise terms  involve $\perp$-derivatives, they must therefore be unrenormalized graphically. This implies that $\gamma$ and $\mu_x$ are unrenormalized graphically.

Since we have already established that
the ratio $\lambda_1/\gamma$ gets no graphical correction, the statement that $\gamma$ itself gets no renormalization then implies that $\lambda_1$ gets no graphical renormalization either.}

\vspace{.2in}

\section{Argument for the irrelevance of ${\alpha\over v_0}\left(u_x+{u^2\over 2v_0}\right)$ }
We use Eq.~(5a) in the main text (MT) to solve for ${\alpha\over v_0}\left(u_x+{u^2\over 2v_0}\right)$ and insert the result into (5b) in the MT to obtain
\beqn
\pp_t \bu_{_\perp}&=&-\vnab_{_\perp} P-\gamma\pp_x\bu_{_\perp}-\lambda_1(\bu_{_\perp} \cdot \vnab_{_\perp})\bu_{_\perp}-{1\over v_0}\left[-\pp_xP-(\gamma+b)\pp_xu_x-\pp_t u_x+f_{_Q}^x+f_{_A}^x\right]\bu_{_\perp}\nonumber\\
&&+\mu_{_\perp}\nabla^2_{_\perp}\bu_{_\perp}+\mu_x\pp_x^2\bu_{_\perp}+\bff_{_Q}^\perp+\bff_{_A}^\perp\,.
\label{Genperp3D1}
\eeqn

Then we can do a preliminary simplification on  (\ref{Genperp3D1}) by neglecting obviously irrelevant terms. For instance, the nonlinear term $\bu_{_\perp}\partial_x u_x$ is irrelevant in comparison to the linear term $\partial_x\bu_{_\perp}$, since both have the same number of $x$-derivative and power in $\bu_{_\perp}$ but the former  has one more power in $u_x$. Following the same reasoning, ${\ctea(}\pp_tu_x{\ctea )}\bu_{_\perp}$ is irrelevant in comparison to $\pp_t\bu_{_\perp}$, and $f_{_{A,Q}}^x\bu_{_\perp}$ are irrelevant in comparison to $\bff_{_{A,Q}}^\perp$. Thus we have a much simplified model:
\beqn
\pp_t \bu_{_\perp}=-\vnab_{_\perp} P-\gamma\pp_x\bu_{_\perp}-\lambda_1(\bu_{_\perp} \cdot \vnab_{_\perp})\bu_{_\perp}+{1\over v_0}\bu_{_\perp}\pp_xP
+\mu_{_\perp}\nabla_{_\perp}^2\bu_{_\perp}+\mu_x\pp_x^2\bu_{_\perp}+\bff_{_Q}^\perp+\bff_{_A}^\perp\,.
\label{Genperp3D2}
\eeqn

Now we show that $\bu_{_\perp}\pp_xP$ is irrelevant. Naively one could argue that this term is irrelevant in comparison to $\vnab_{_\perp} P$, since $\bu_{_\perp}\pp_xP$ has one more power in $\bu_{_\perp}$ and $x$-derivative power counts less than $\perp$-derivative as indicated by the linear theory. However, this argument is {\it not} sufficient. This is because in the equation of motion for $\bu_{_T}$, $\vnab_{_\perp} P$ is absent while $\bu_{_\perp}\pp_xP$ (more precisely, $\bu_{_T}\pp_xP$)  remains. To see that $\bu_{_\perp}\pp_xP$ is irrelevant we have to take a winding path. We take divergence on both sides of the equality in (\ref{Genperp3D2}).
We get
\beqn
\vnab_{_\perp}^2 P=\vnab_{_\perp}\cdot\left[-\pp_t \bu_{_\perp}-\gamma\pp_x\bu_{_\perp}
-\lambda_1(\bu_{_\perp} \cdot \vnab_{_\perp})\bu_{_\perp}
+\mu_{_\perp}\nabla_{_\perp}^2\bu_{_\perp}+\mu_x\pp_x^2\bu_{_\perp}
+\bff_{_Q}^\perp+\bff_{_A}^\perp\right]\,,
\label{Pressure1}
\eeqn
where we have neglected  $\nabla_{_\perp}\cdot\left(\bu_{_\perp}\pp_xP\right)$, which is obviously irrelevant in comparison to $\vnab_{_\perp}^2 P$. This shows that $P$ power counts like
\beqn
{\vnab_{_\perp}\over \vnab^2_{_\perp}}\cdot\left[-\pp_t \bu_{_\perp}-\gamma\pp_x\bu_{_\perp}
-\lambda_1(\bu_{_\perp} \cdot \vnab_{_\perp})\bu_{_\perp}
+\mu_{_\perp}\nabla^2_{_\perp}\bu_{_\perp}+\mu_x\pp_x^2\bu_{_\perp}
+\bff_{_Q}^\perp+\bff_{_A}^\perp\right]\,,
\label{Pressure2}
\eeqn
and hence $\bu_{_\perp}\pp_xP$ power counts like
\beqn
\bu_{_\perp}\pp_x\left\{{\vnab_{_\perp}\over \vnab^2_{_\perp}}\cdot\left[-\pp_t \bu_{_\perp}-\gamma\pp_x\bu_{_\perp}
-\lambda_1(\bu_{_\perp} \cdot \vnab_{_\perp})\bu_{_\perp}
+\mu_{_\perp}\nabla^2_{_\perp}\bu_{_\perp}+\mu_x\pp_x^2\bu_{_\perp}
+\bff_{_Q}^\perp+\bff_{_A}^\perp\right]\right\}\,.
\label{Pressure3}
\eeqn
Since $x$-derivative power counts less than $\perp$-derivative, (\ref{Pressure3}) power counts less than
\beqn
\bu_{_\perp}\left[-\pp_t \bu_{_\perp}-\gamma\pp_x\bu_{_\perp}
-\lambda_1(\bu_{_\perp} \cdot \vnab_{_\perp})\bu_{_\perp}
+\mu_{_\perp}\nabla^2_{_\perp}\bu_{_\perp}+\mu_x\pp_x^2\bu_{_\perp}
+\bff_{_Q}^\perp+\bff_{_A}^\perp\right]\,.
\label{Pressure4}
\eeqn
Therefore, all the above terms are irrelevant in comparison to the ones  in (\ref{Genperp3D2}). For instance, the first term in (\ref{Pressure4}) is irrelevant versus $\pp_t \bu_{_\perp}$ in (\ref{Genperp3D2}) since they have same number and same type of derivatives while the former has one more power in $\bu_{_\perp}$. The same argument applies to other terms in (\ref{Pressure4}). Therefore, we conclude that $\bu_{_\perp}\pp_xP$ in (\ref{Genperp3D2}) is irrelevant.

Neglecting $\bu_{_\perp}\pp_xP$ in (\ref{Genperp3D2}), we obtain
\beqn
\pp_t \bu_{_\perp}=-\vnab_{_\perp} P-\gamma\pp_x\bu_{_\perp}-\lambda_1(\bu_{_\perp} \cdot \vnab_{_\perp})\bu_{_\perp}+\mu_{_\perp}\nabla_{_\perp}^2\bu_{_\perp}+\mu_x\pp_x^2\bu_{_\perp}+
\bff_{_Q}^\perp+\bff_{_A}^\perp\,,
\label{}
\eeqn
which is exactly Eq.~(12) in the MT.

\vspace{.2in}

\section{Detailed derivation of $\langle \bu(\br, t)\cdot\bu(\mathbf 0, 0)\rangle$}
By changing the variables of integration  to
\beq
{\bk}_{_\perp} \equiv \bq_{_\perp} (r_{_\perp}\Lambda)\,,~~~
k_{_x} \equiv q_{_x} (r_{_\perp}\Lambda)^{\zeta},
\eeq
Eq.~(20b) in the MT can be written as
\beqn
C_{_{Q}}(\br)=r_{_\perp}^{2\chi}G_{_{Q}}\left(|x|\over r_{_\perp}^{\zeta}\right)\,,
\eeqn
where
\bew
\beqn
G_{_Q}\left(|x|\over r_{_\perp}^{\zeta}\right)\equiv
{\Lambda^{2\chi}\over (2\pi)^d}\int \dd^d k
\left[{2(d-2) D_{_Q}\over {\gamma^2k_x^2+\left[\mu_{_{\perp 0}}k_{_\perp}^2(k_{_\perp}/\Lambda)^{-\eta_{_\perp}}f_{\mu_{_\perp}}\left({k_x/\Lambda'\over (k_{_\perp}/\Lambda)^{\zeta}}\right)\right]^2}}\right]\ee^{\ii\left[{xk_x\over (r_{_\perp}\Lambda)^{\zeta}}
+{\bk _{_\perp}\cdot \br_{_\perp}\over r_{_\perp}\Lambda}\right]}\,.
\eeqn
\ew

Inserting Eq.~(23) in the MT into Eq.~(20a) in the and  introducing $\omega'=\omega-\gamma q_x$, we obtain
\bew
\beqn
C_{_{A}}(\br, t)
=\int \frac{\dd\omega'\dd^d q}{(2\pi)^{d+1}}
\left[{2(d-2)D_{_A}\over \left(\omega'\right)^2+\left[\mu_xq_x^2+
\mu_{_{\perp 0}}\left(q_{_x}\over\Lambda'\right)^{-{\eta_{_\perp}\over\zeta}}q_{_\perp}^2\right]^2}\right]
\ee^{\ii(\bq\cdot\br-\omega t)}\,.
\eeqn
\ew
Further changing the variables of integration  to
\beq
{\bk}_{_\perp} \equiv \bq_{_\perp} (r_{_\perp}\Lambda)\,,~~~
k_{_x} \equiv q_{_x} (r_{_\perp}\Lambda)^{\zeta'}\,,~~~\Omega \equiv \omega' (r_{_\perp}\Lambda)^{z'}
\eeq
we obtain
\beqn
C_{_{A}}(\br, t)=r_{_\perp}^{2\chi'}G_{_{A}}\left({|x-\gamma_0t|\over r_{_\perp}^{\zeta'}},{|t|\over r_{_\perp}^{z'}}\right)\,,
\label{Scaling function sm1}
\eeqn
where
\bew
\beqn
G_{_A}\left({|x-\gamma_0t|\over r_{_\perp}^{\zeta'}},{|t|\over r_{_\perp}^{z'}}\right)
\equiv{2(d-2)D_{_A}\Lambda^{2\chi'}\over (2\pi)^{d+1}}\int \dd\Omega\dd^dk\,
{{{\exp\left\{-\ii\left[{\Omega t\over (r_{_\perp}\Lambda)^{z'}}-{\bk_{_\perp}\cdot\br_{_\perp}\over r_{_\perp}\Lambda}-{(x-\gamma t)k_x\over (r_{_\perp}\Lambda)^{\zeta'}}\right]\right\}}\over\Omega^2+\left[\mu_xk_x^2+
	\mu_{_{\perp 0}}k_{_\perp}^2(k_x/\Lambda')^{-{\eta_{_\perp}\over\zeta}}\right]^2}}\,,
\label{Scaling function sm}
\eeqn
\ew
\beqn
&&\chi'={1\over 2}\left[z'-\zeta'-(d-1)\right]\,,\\
&&z'=2\zeta'\,,\\
&&\zeta'={2\zeta\over 2\zeta+\eta_{_\perp}}\,.
\eeqn
Inserting the values of $\eta_{_\perp}$ and $\zeta$ from Eq.~(16) in the MT into the above expressions we get the values of $z'$, $\zeta'$, and $\chi'$ quoted in the main text (2).

\begin{figure}
\begin{center}
	\includegraphics[scale=.65]{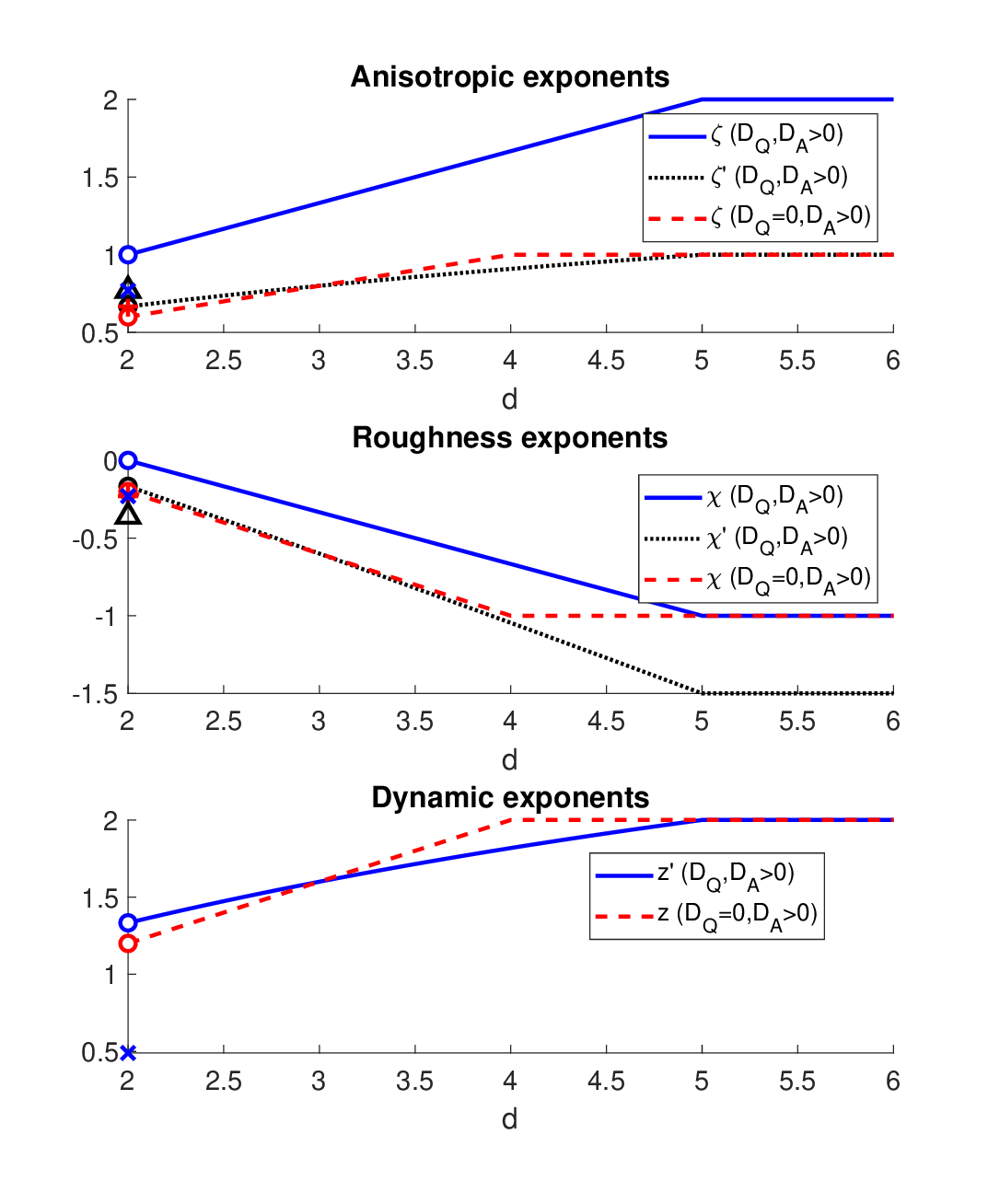}
\end{center}
\caption{The scaling exponents plotted versus  spatial dimension $d$  in incompressible polar active fluids with quenched disorder ($D_{_Q}> 0$, $D_{_A}> 0$, solid blue line and dotted black line) and with annealed disorder only ($D_{_Q}=0$, $D_{_A}> 0$, red dashed line). The scaling exponents are the anisotropy exponent $\zeta$ ($\zeta'$ denotes the anisotropy exponent for the annealed part of the correlations in the presence of quenched disorder), the roughness exponent $\chi$ ($\chi'$ denotes the roughness exponent for the annealed part of the correlations in the presence of quenched disorder), and the dynamic exponent $z$ ($z'$ denotes the dynamic exponent for the annealed part of the correlations in the presence of quenched disorder).
	In both the cases with quenched disorder and annealed disorder only, the $d>2$ results do not converge to the $d=2$ results (blue crosses and black triangles for the case with quenched  disorder \cite{us} and red pluses for the case with annealed disorder only \cite{chen_natcomm16,chen_njp18}).	{Note that the dynamic exponent for the case with annealed disorder only in $d=2$ remains unknown.}	The upper critical dimensions are $5$ and $4$ for the cases with quenched disorder and annealed disorder only, respectively.
}
\label{fig:exponents}
\end{figure}

In Fig.~\ref{fig:exponents}, we plot the scaling exponents as a function of spatial dimension.

{
\section{Calculation of the exponent $\theta$}
We begin with the scaling form (\ref{Scaling function sm}) for the annealed part of the correlation function, which
contains all of the time dependence.  If we sit at fixed, large {$\br_\perp$  (i.e., $x=0$, large $\br_\perp$)}, at long times the two
arguments of the scaling function in equation (\ref{Scaling function sm})  go like $|t|/r_\perp^{z'}$ and $|t|/
r_\perp^{\zeta'}$, respectively. The correlation function will be controlled by the larger of these
two, which, for large $r_\perp$, is $|\gamma t|/r_\perp^{\zeta'}$, since $\zeta'<z'$. Therefore, for large time,
\beq
G_{_A}\left({|\gamma t|\over r_{_\perp}^{\zeta'}},{|t|\over r_{_\perp}^{z'}}\right)
\approx G_{_A}\left({|\gamma t|\over r_{_\perp}^{\zeta'}},0\right)\equiv  F_{_A}\left({|\gamma t|\over r_{_\perp}^{\zeta'}}\right)\,.
\label{Scaling function theta}
\eeq

We expect that, at sufficiently large time $t$ {(i.e., $|t|\gg r_{_\perp}^{\zeta'}$)},
the correlation function (\ref{Scaling function sm1})  should become independent of {$\br_{_\perp}$}. This implies that the scaling function $F_{_A}$ in (\ref{Scaling function theta}) must scale like $r_\perp^{-{2\chi'}}$, to cancel off the $r_\perp^{{2\chi'}}$ prefactor in (\ref{Scaling function sm1}). But since the scaling function  $F_{_A}(u)$ in (\ref{Scaling function theta}) depends only on the scaling combination $u\equiv {|\gamma t|\over r_{_\perp}^{\zeta'}}$, this can only be achieved if
\beq
F_{_A}(u)\propto u^\theta
\label{FAcond}
\eeq
with
\beq
\theta=\frac{2\chi'}{\zeta'} =-\left[{d^2+4d-9\over 2(d+1)}\right]\,.
\label{thetasm}
\eeq
Requiring this scaling of $F_{_A}(u)$ clearly implies, from (\ref{Scaling function sm1}), that
\beq
C_{_{A}}(x=0,\br_{_\perp}={\mathbf 0}, t)\propto |t|^\theta\,,
\label{CAfin}
\eeq
at large time, with $\theta$ given by (\ref{thetasm}), which is the result quoted in the main text.

}
\end{document}